\documentclass{egpubl}
\usepackage{eg2021}
 
\ConferencePaper        % uncomment for (final) Conference Paper
\usepackage[T1]{fontenc}
\usepackage{dfadobe}  
\usepackage{amsmath}
\usepackage{amssymb}

\biberVersion
\BibtexOrBiblatex
\usepackage[backend=biber,bibstyle=EG,citestyle=alphabetic,backref=true]{biblatex} 
\electronicVersion
\PrintedOrElectronic

\ifpdf \usepackage[pdftex]{graphicx} \pdfcompresslevel=9
\else \usepackage[dvips]{graphicx} \fi

\usepackage{egweblnk} 
\usepackage[table]{xcolor}

\newcommand{\sarahchange}[1]{\textcolor[rgb]{0 0 0}{#1}}
\newcommand{\risachange}[1]{\textcolor[rgb]{0 0 0}{#1}}

\newcommand{\ra}[1]{\renewcommand{\arraystretch}{#1}}
\definecolor{lightbluishgrey}{rgb}{0.76471,0.84824,0.91647}
\usepackage{tabularx}
\usepackage{booktabs}

\newcommand{\refequ}[1] {Eq.~\ref{equ:#1}}

\newcommand{\reffig}[1] {Figure~\ref{fig:#1}}
\newcommand{\reffigs}[2] {Figures~\ref{fig:#1},\ref{fig:#2}}

\newcommand{\reftab}[1] {Table~\ref{tab:#1}}
\newcommand{\refsec}[1] {Sec.~\ref{sec:#1}}
\newcommand{\refapp}[1] {App.~\ref{app:#1}}

\renewcommand{\a}{\mathbf{a}}
\renewcommand{\t}{\mathbf{t}}
\newcommand{\that}{\hat{\mathbf{t}}}
\renewcommand{\c}{\mathbf{c}}
\newcommand{\e}{\mathbf{e}}
\newcommand{\f}{\mathbf{f}}
\newcommand{\m}{\mathbf{m}}

\renewcommand{\d}{\mathbf{d}}
\newcommand{\g}{\mathbf{g}}
\newcommand{\x}{\mathbf{x}}
\newcommand{\xbar}{\overline{\mathbf{x}}}
\newcommand{\N}{\mathbf{N}}
\newcommand{\n}{\mathbf{n}}
\renewcommand{\S}{\mathbf{S}}

\newcommand{\D}{\mathbf{D}}
\newcommand{\GS}{\mathcal{G}}
\newcommand{\C}{\mathbf{C}}

\newcommand{\R}{\mathbb{R}}
\newcommand{\Zero}{\mathbf{0}}
\newcommand{\q}{\mathbf{q}}
\newcommand{\Q}{\mathbf{Q}}

\newcommand{\pluseq}{\mathrel{+}=}
\newcommand{\minuseq}{\mathrel{-}=}

\usepackage{wrapfig}

\title[Levitating Rigid Objects with Hidden Rods and Wires]{\vspace*{-0.25cm}Levitating Rigid Objects with Hidden Rods and Wires\vspace*{-0.5cm}}
\author[Kushner et al.]
{\parbox
    {\textwidth}
        {\centering 
            Sarah Kushner, Risa Ulinski, Karan Singh, David I.W. Levin, Alec Jacobson
        }
        \\
\vspace*{-0.25cm}
{\parbox
    {\textwidth}
        {\centering 
            University of Toronto
        }
    }
}
\begin{document}

\teaser{
  \vspace*{-0.5cm}
  \includegraphics[width=\linewidth]{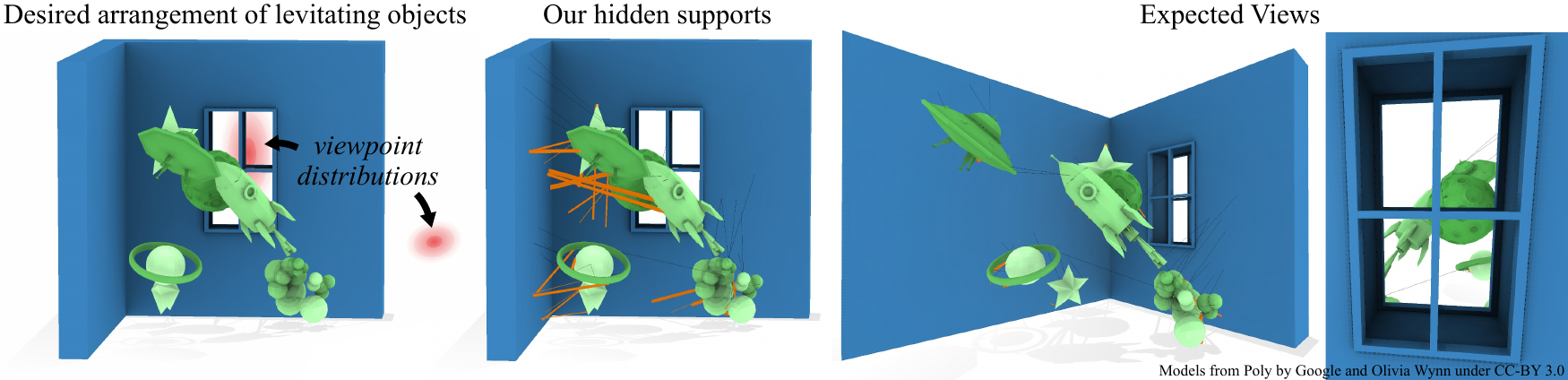}
 \centering

  \caption{\label{fig:teaser} 
  Our optimization finds hidden supports to hold rigid objects (green) in their
  locations despite gravity. Rods (orange) resist tension,
  compression and bending, while wires (black) resist tension. Supports connect
  between objects or to the input support surface (blue). Rods are \emph{hidden}
  behind occlusions in the scene for a possibly disconnected distribution of
  viewpoints (red) provided by the user.
Here, a
collection of space-themed objects seemingly hover in the corner of a room. The supporting
truss is hidden from the front \emph{and} through the window. }
}

\maketitle

% \begin{CCSXML}
%     <ccs2012>
%     <concept>
%     <concept_id>10010147.10010371.10010352.10010381</concept_id>
%     <concept_desc>Computing methodologies~Collision detection</concept_desc>
%     <concept_significance>300</concept_significance>
%     </concept>
%     <concept>
%     <concept_id>10010583.10010588.10010559</concept_id>
%     <concept_desc>Hardware~Sensors and actuators</concept_desc>
%     <concept_significance>300</concept_significance>
%     </concept>
%     <concept>
%     <concept_id>10010583.10010584.10010587</concept_id>
%     <concept_desc>Hardware~PCB design and layout</concept_desc>
%     <concept_significance>100</concept_significance>
%     </concept>
%     </ccs2012>
% \end{CCSXML}

% \ccsdesc[300]{Computing methodologies~Collision detection}
% \ccsdesc[300]{Hardware~Sensors and actuators}
% \ccsdesc[100]{Hardware~PCB design and layout}

% \printccsdesc 
\begin{abstract}
We propose a novel algorithm to efficiently generate hidden structures to
support arrangements of floating rigid objects.
Our optimization finds a small set of rods and wires between objects and each
other or a supporting surface (e.g., wall or ceiling) that hold all objects in
force and torque equilibrium.
Our objective function includes a sparsity inducing total volume term and a
linear visibility term based on efficiently pre-computed Monte-Carlo
integration, to encourage solutions that are \emph{as-hidden-as-possible}.
The resulting optimization is convex and the global optimum can be efficiently
recovered via a linear program.
Our representation allows for a user-controllable mixture of tension-,
compression-, and shear-resistant rods or tension-only wires.
We explore applications to theatre set design, museum exhibit curation, and
other artistic endeavours.
\end{abstract}

\begin{wrapfigure}[12]{r}{0.4\linewidth}
    \vspace*{-1\intextsep}
    \hspace*{-0.5\columnsep}
    \begin{minipage}[b]{1.1\linewidth}
    \includegraphics[width=\linewidth, trim={0mm 0mm 0mm 0mm}]{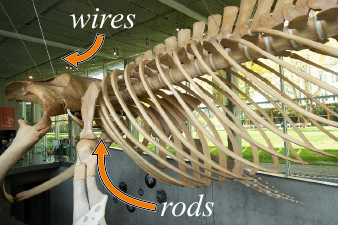}
  \caption{The skeleton of a blue whale levitates with the support of wires from
  above and internal rods.
  \label{fig:ubc-whale}}
    \end{minipage}
\end{wrapfigure}
\section{Introduction}
\begin{figure}[!b]
  \includegraphics[width=\linewidth]{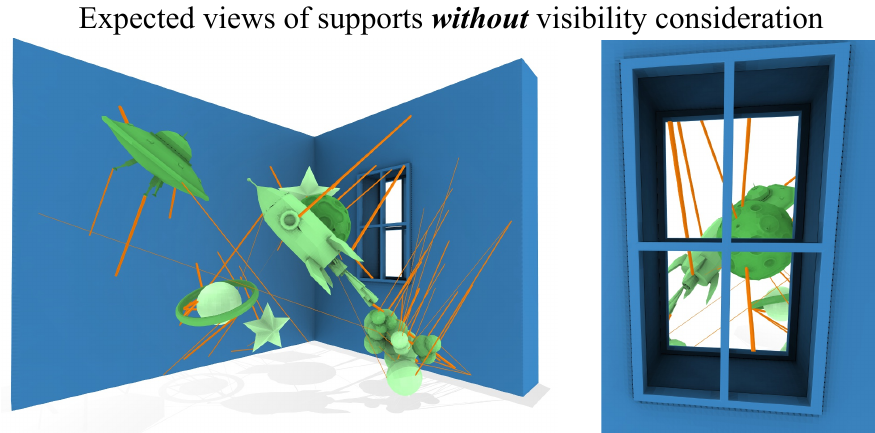}
 \centering
  \caption{
\label{fig:rocket-visible}
  Without our visibility term, optimal rods may be an unsightly distraction.
  \vspace*{-1cm}
  }
\end{figure}

\begin{figure}
  \includegraphics[width=\linewidth]{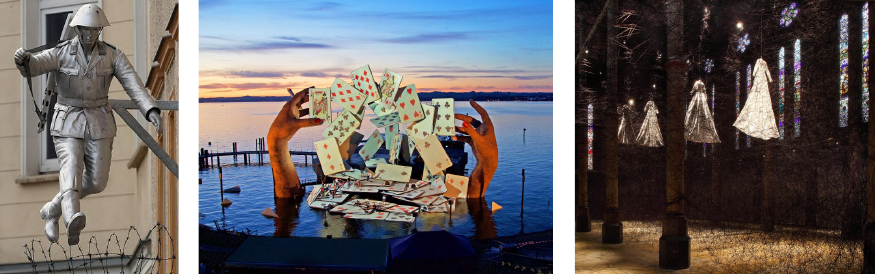}
 \centering
  \caption{
\label{fig:art}
  Levitating objects have inspired such artworks as 
  a sculpture of border guard Conrad Schumann jumping (left), 
  an enormous stage display of playing cards by Es Devlin for the Bregenz Festival (middle),
  and Chiharu Shiota's installation where white dresses float overhead (right).
  \vspace*{-.75cm}
  }
\end{figure}

%
% Narrow in on the topic
%
Levitating objects are visually compelling and commonly found in artistic sculptures,
film and theatre set design, promotional displays, and museum exhibits (see
\reffig{ubc-whale} and \reffig{art}).
This effect is especially impressive if the support structure can be hidden from
the observer, removing its unsightly distraction and perhaps even giving the
impression that the objects in the arrangement are magically floating in space
(see \reffig{teaser}).
%
% Ancient sculptures and logs/cherubs?
%
% Dig a hole
Achieving this is a non-trivial task.
Physical stability requires a balance of force and torque for each rigid component of
the scene.
This is readily achieved using many strong, thick struts, but their geometry and scene 
placement is likely to compete for visual attention with scene objects, or worse, visually obscure objects in the scene (see
\reffig{rocket-visible}).
Hiding these supports by removing or thinning too many struts, on the other hand, will sacrifice
physical stability.
Thin wires can sometimes be used to hang objects, but wires only resist tension
so they alone can not handle situations that are not supported purely from above.

% Try to keep this on the first page
In this paper, we propose modeling the problem of hidden support structure
generation for \emph{levitating objects} as a form of topology optimization.
We present a novel convex optimization based on the well-established
\emph{ground structure method} from architecture and engineering.
The input to our method is an arrangement of objects in their desired locations
and orientations and the distribution of views from which the scene will likely
be observed.
Our output is a collection of \emph{rods} and \emph{wires}, described by their
required thicknesses and attachment points on the input rigid objects, and the
supporting structural element (e.g., wall or ceiling).
Our rods model tension, compression and bending resistant materials (e.g.,
wooden dowel rods or steel beams).
Our wires model tension only (e.g., fishing line or steel cables).

Unlike Computer Graphics or Virtual Reality where physical laws can be bent or
broken, support structures in real scenes are only meaningful if physically
valid.
Therefore, we enforce physical validity in our optimization as a hard
constraint: namely that the rigid objects should achieve force and torque
equilibrium and that stresses on rods and wires do not exceed material-dependent
yield limits.
For ease of assembly, cost of manufacturing, and visibility considerations, we prefer
support structures composed of a small number of thin, less visible supports. %See \reffig{}.
We model these criteria with a sparsity-inducing cost function defined as a sum
over a densely connected graph of edges (i.e., the \emph{ground
structure}).

Treating the cross-sectional area of each edge as the primary optimization variable, the
traditional ground structure method optimizes the total volume (linear in the
areas since lengths are predetermined) and enforces force balance at point loads,
by measuring linearized axial tension and compression forces from each rod,
subject to yield limits, expressed as linear inequalities in the unknown cross-section areas and
axial stresses of the rods.
The result is a linear program whose solution --- like many $L_1$ or Lasso
problems --- is sparse (most areas are exactly zero), and often agrees exactly
with the NP-hard selection problem (picking the smallest valid subset of edges).

We augment the traditional ground structure method to support embedded rigid
objects (via linear static equilibrium equations) and account for bending
resistance of rods (via a simple linear shearing model derived from
proportionality assumptions).
We introduce a visibility objective function that is also linear in the unknown
edge areas and relies on efficient Monte-Carlo based precomputation.
Thus, the optimization remains a (convex) linear program and solutions can be
extracted efficiently (in usually less than a minute).

Our experiments satisfyingly confirm that under many conditions structurally
valid supports are lurking just out of sight: the space of physically valid
supports is vast and finding a completely occluded arrangement is often
possible.
We demonstrate the effectiveness of our method across a wide variety of test
scenes and prototypical use cases.

\begin{figure*}
    \includegraphics[width=\linewidth]{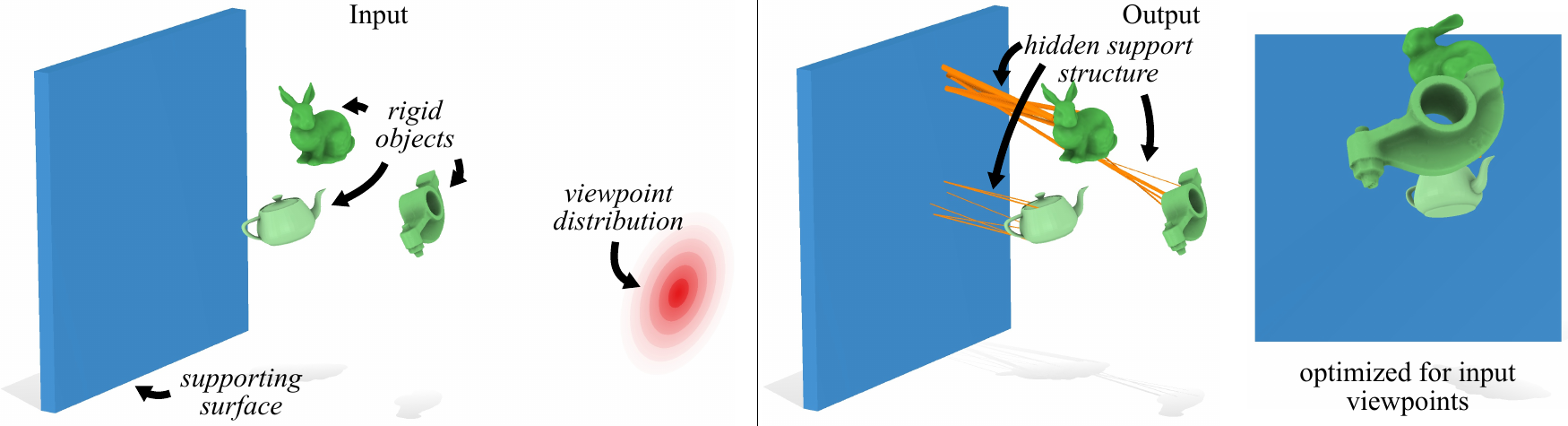}
   \centering
    \caption{
      \label{fig:bunny-teapot-rocker-arm}
      The input to our method is a scene composed of many levitating rigid
      objects. The output of our method is a collection of rods tucked away 
      behind object occlusions, holding each
      object in force and torque equilibrium under gravity. 
    }
    \vspace{-6mm}
\end{figure*}

\section{Related Work}
\label{sec:related}
Our work sits within the larger literature of computational fabrication,
construction and assembly.
These subfields are rich and vast, so we focus on previous works most similar in
methodology or application.

Previous algorithms exist to make objects
stand~\cite{PrevostWLS13,VougaHWP12}, spin \cite{Baecher14} or hang from
wires~\cite{Mobiles}.
These works modify the input objects by redistributing mass or changing their
shape to achieve the desired goal.
In this paper, we explore a complementary contract with the user --- how to
anchor objects in the environment \emph{without} changing the objects
themselves.
We do not assume that objects were fabricated in a particular manner (e.g., 3D
printing).

Our approach may be categorized with other structural optimizations for a
prescribed static load scenario (i.e., ignoring inertial forces).
Recent works increase the stability of fragile objects by adding new
structural elements~\cite{ZhouPZ13,StavaVBCM12,Coros16}.
For example, Stava et al. add struts to 3D printed objects one-by-one as part of
a large optimization loop and use a volumetric simulation as validation. Their
strut selection includes an ambient occlusion visibility term, but they do not
consider the problem of selecting an optimal set of supports for rigid objects
under prescribed viewing conditions.
Other methods have considered the interactive design of rod-structures
~\cite{PerezTCBCSO15,KovacsSWCMMYBSK17,ChidambaramZSER19,Jacobson19} with varying degrees of
physical feasibility checking or optimization in the system.

%For our problem, we take the alternate approach -- beginning with an overly
%prescribed structure, one which contains more structural elements than
%necessary, we prune unneeded elements away via an optimization algorithm. This
%is called Topology optimization and it comes in two flavors: volumetric and
%truss-based.  In the volumetric edition, the volume of the structure is
%discretized using space filling primitives. The optimization prescribes an
%amount of material to each primitive, leading to some areas of the volume being
%left empty. A common objective is to minimize the compliance of the structure
%under the constraint that it remain in static equilibrium. These algorithms have
%seen wide adoption, however discretizing the volume can be onerous, especially
%in cases where one wishes to pass loads over long distances. Such cases are more
%amenable to the use of structural members like beams, trusses and cables.

We model the problem of hiding support structures as a form of topology
optimization \cite{liu2018}.
The general idea of topology optimization is to prune away material from the
volume around the input objects or load conditions.
The resulting geometries typically have interesting topologies/connectivities
that would have been difficult to determine \emph{a priori}.
Methods that determine the material occupancy of each voxel in a dense grid are
well suited for 3D printing and milling (e.g., \cite{DWuDW16}), but will in
general produce geometries composed curved and varying thickness elements.
Our method instead belongs to the class of \emph{ground structure} methods
\cite{dorn1964automatic},
which output a discrete collection of (straight) elements from an initial
over-connected graph of candidates (see \reffig{bunny-ground-structure}).
Methodologically we follow most closely the stress-based formulation of Zegard
et al.~\cite{zegard2015grand3}, and utilize the thesis of Freund
\cite{freund2004} as a reference.
Ground structure-like methods have
been applied for designing everything from buildings~\cite{zegard2020} 
\sarahchange{and glass shell structures ~\cite{DBLP:journals/cgf/LacconeMFCP20}} to construction
supports~\cite{PanozzoS14} to 3D printable models~\cite{WangWYLTTDCL13,
JiangTSW17,HuangZHSLYL16} to cable-driven automata~\cite{MegaroKSLMGTB17}.
The standard ground structure method considers only axial
forces. These methods have been applied for rigid structural elements and
adapted to special cases like tensegrities~\cite{PietroniTVPC17,ConnellyW96}.  

We use a ground structure approach to model the novel problem of
creating hidden structural supports from complex viewpoint distributions.
Crucially, our method supports structural elements that resist compression,
tension and bending forces, as well as wires, without resorting to the nonlinear
constitutive models or volumetric meshing of prior
work~\cite{StavaVBCM12,PerezTCBCSO15,HuangZHSLYL16}.
Our method trivially couples the structure to the rigid objects it supports,
correctly accounting for both linear forces and torques, without resorting to
displacement-based mechanical formulations (e.g., \cite{PietroniTVPC17}).
This allows us to formulate our problem as a linear program which can be
solved efficiently.
% \sarahchange{and will give a sparse solution} ~\cite{candes2008enhancing,feng2013complementarity}.
%
% \risachange{We ensure a the resulting ground structure is sparse using $L_1$ regularization as 
% described in ~\cite{candes2008enhancing,feng2013complementarity}.} 
% \sarahchange{Our formulation also induces a sparse solution thanks to $L_1$ regularization
%  ~\cite{candes2008enhancing,feng2013complementarity}.}

We draw inspiration from algorithms for appearance-driven optimization. For
instance, Schuller et al. introduce the problem of generating appearance
mimicking surfaces from a specified
viewpoint~\cite{Schuller:2014:AS:2661229.2661267}.
Several works seek to create 3D shapes that take the form of a set of 2D shapes
from corresponding viewpoints or cast the 2D image under certain lighting
conditions~\cite{Mitra:2009:SA:1618452.1618502,Hsiao:2018:MWA:3272127.3275070,Schwartzburg14}.
Others use viewpoints to create optimal perceptual experiences, for example in
3D printing support structures~\cite{ZhangLPWW15} or in skyscraper
design~\cite{DoraiswamyFLVWW15}.

\begin{wrapfigure}[9]{r}{1.45in}
  \includegraphics[width=\linewidth,trim={6mm 0mm 0mm 10mm}]{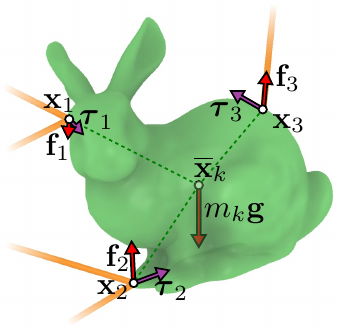} 
  \label{fig:bunny-equilibrium}
\end{wrapfigure} 

\section{Method}
\label{sec:method}
The input to our method is a scene comprised of $K$ rigid objects oriented and
positioned in space, a fixed support surface (e.g., wall or ceiling), and a
distribution of viewpoints (e.g., discrete set of positions or sample-able
probability density function defined on a surface) 
% \review{why surfaces? confusion about the viewing direction. maybe we should explain that the directions come from \textbf{ex} earlier? maybe it would be useful to have an example where the viewing surface is not a plane? i guess we decided that these distributions were good enough to approximate where people would be standing to look at our examples}.
%
The output of our method is a \emph{supporting structure} composed of a small
set of \emph{rods} and \emph{wires} connecting rigid objects to each other or
the supporting surface. Our method ensures that this structure holds the input
objects in their prescribed positions and orientations, counter-balancing the
force these objects experience due to gravity. Our method optimizes the size and
placement of the structure to minimize its overall volume and its visibility
with respect to the input viewpoint distribution (see
\reffig{bunny-teapot-rocker-arm}). 
Before describing our optimization, we define our physical model and how we
measure visibility.

\subsection{Rigid Body Equilibrium}
\label{sec:equilibrium}
The rigid objects in our scenes experience forces from gravity and at the points
of attachment to the supporting structure.
To hold a rigid body at rest, we must maintain force and torque
equilibrium:
\begin{align}
  \sum\limits_{i \in V_k} \f_i  & = m_k \g, \\
  \sum\limits_{i \in V_k} \underbrace{(\x_i-\xbar_k) \times
  \f_i}_{\boldsymbol{\tau}_i} & = \mathbf{0},
\end{align}
where $m_k$, $\xbar_k$, and $V_k$ are the mass, center of mass, and set of
attachment points of the $k$th object, respectively, and
$\x_i,\f_i,\boldsymbol{\tau}_i$ are the 3D position of the $i$th attachment
point and corresponding force and torque vectors, respectively.

\subsection{Rods}
\label{sec:rods}
We assume our support structure undergoes negligible displacement, affording a
linearization of the internal forces at play.
For stiff rods, we follow the linearized tension and compression model of
\cite{zegard2015grand3,freund2004}, which introduces a \emph{signed} scalar value per
rod $c_{ij} \in \R$ with units Newtons describing the force in the axial direction parallel to the rod.
Assigning an arbitrary direction to the rod $ij$ between endpoint positions
$\x_i$ and $\x_j$, then the axial force contribution at endpoints $i$ and $j$
are the product of this scalar $c_{ij}$ by the rod's tangent unit direction
$\that_{ij} = (\x_i-\x_j)/\|\x_i-\x_j\|$:
\begin{align}
  \f_i \pluseq   c_{ij} \that_{ij} \quad \text{ and } \quad
  \f_j \minuseq  c_{ij} \that_{ij}.
\end{align}

\begin{figure}
  \includegraphics[width=\linewidth]{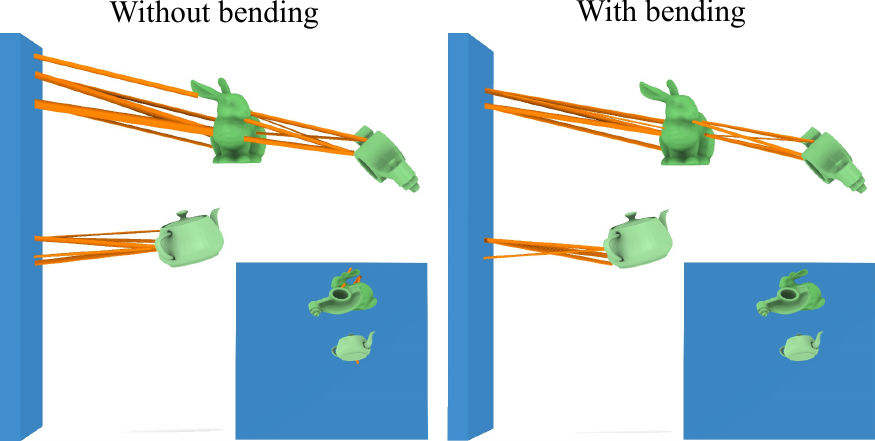}
  \caption{\label{fig:benefit-bending}
  \sarahchange{
  The addition of our linearized bending term yields sparser, less visible support 
  structures by more accurately modelling the strength of the rods. Rods can also connect
  between objects rather than just to the support surface.}}
  \vspace*{-2mm}
\end{figure}

Previous methods (e.g., \cite{zegard2015grand3,freund2004}) rely solely on
tension and compression and ignoring the rods' resistance to \emph{bending}.
This is a reasonable assumption in architecture where loads are
large relative to the rod's bending strength. \sarahchange{Ignoring
bending requires that the rods are thicker and thus more visible (see \reffig{benefit-bending})}.
This is at odds with the intuition that light loads can be held up with a single
bending-resistant rod. In reality, a single rod with finite thickness can apply
a \emph{distribution} of forces over its non-zero area contact surface. Since
the force is applied at more than one point, torque balance is also possible.
Unfortunately, a volumetric rod model couples the unknown rod diameters and
forces non-linearly.

\begin{figure}[t!]
  % \vspace*{-0.5cm}
  \includegraphics[width=\linewidth]{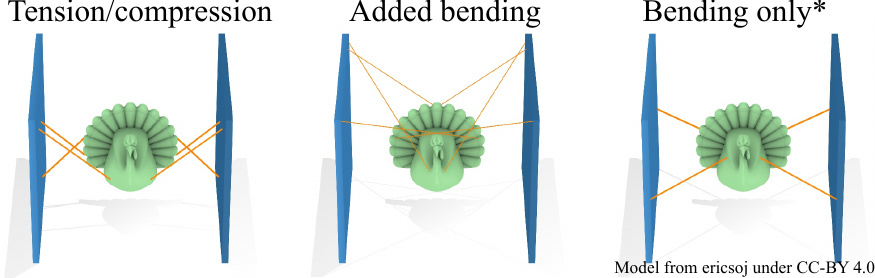}
  \caption{\label{fig:turkey-bending}
  An enormous turkey levitates between two buildings using tension and
  compression resistant rods (left). Adding bending resistance affords a less
  voluminous solution (middle). Restricting the ground structure to only include
  edges perfectly intersecting the center of mass $(^*)$ admits a bending only
  solution (right).
  }
  \vspace*{-6mm}
\end{figure}

To maintain the linearity of our system but also account for bending,
we introduce a linearized shearing model to account for resistance in the normal
direction.(see \reffig{turkey-bending}).
% \sarah{go from bending to shearing here.}
%
%
% \sarah{note: but there is some intuition about the force being proportional to the length of
% the rod from bending moments}
% \alec{not sure where/how/if to mention that. Since the shear term no longer
% explicitly involves the length it's not conspicuous...}
%
For each
rod $ij$, we introduce an arbitrary orthonormal basis $\N_{ij} \in \R^{3\times 2}$
for the 2D space orthogonal to the axial direction. We introduce a two
dimensional parameter $\q_{ij}  \in \R^2$ with units Newtons describing
the force on the rod in the two normal basis directions.
Shear force contributions are equal and opposite at either end of each rod:
\begin{equation}
  \f_i \pluseq \N_{ij} \q_{ij}
  \quad \text{ and } \quad
  \f_j \minuseq \N_{ij} \q_{ij}.
\end{equation}
\begin{wrapfigure}[7]{r}{1.2in}
  % \vspace{-0.5cm}
  \includegraphics[width=\linewidth,trim={4mm 0mm 0mm 3mm}]{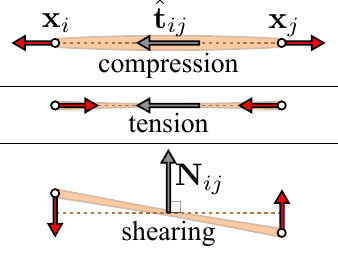} 
  \label{fig:rod-forces}
\end{wrapfigure} 
Following previous methods \cite{zegard2015grand3,freund2004}, we model failure
catastrophically. If the stress due to tension, compression or bending exceeds a
material-dependent fixed threshold we declare that the rod has exploded (or at
least moved too much) and is no longer feasible. These yield stresses can be
prescribed for each rod $ij$ and can be related directly to the non-negative rod
cross-sectional area $a_{ij} \in \R_{\ge 0}$ and the force parameters introduced
above. Namely, we require the following \emph{convex} inequalities to hold:
\begin{align}
  -\sigma^t_{ij} a_{ij} \leq c_{ij} \leq \sigma^c_{ij} a_{ij} \quad \text{and}
  \quad
  \label{equ:yieldb}
  \|\q_{ij}\| \leq \sigma^s_{ij} a_{ij},
\end{align}
where $\sigma^t_{ij},\sigma^c_{ij},\sigma^s_{ij}$ are the tension, compression,
and shearing stress thresholds, respectively. For common rod materials, we find
that $\sigma^t_{ij} \approx \sigma^c_{ij} >> \sigma^s_{ij}$. Although $\sigma^t$
and $\sigma^c$ values for specific materials (e.g., pine wood) can be found in
reference books, in our experience all of these parameters should be empirically
estimated, especially when working with low-end materials from the hardware
store.

\begin{wrapfigure}[16]{r}{1.13in}
  \includegraphics[width=\linewidth,trim={6mm 0mm 0mm 10mm}]{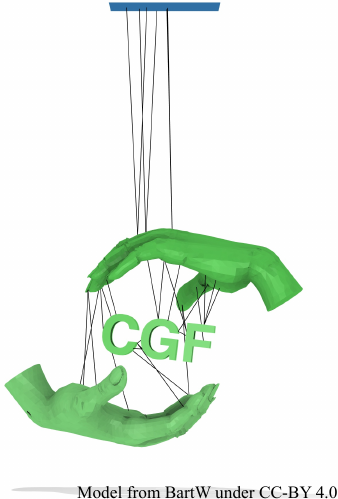} 
    \caption{\label{fig:cupped-sgp} 
    Wire-only solutions require support from above the arrangement's center of
    mass.
  }
\end{wrapfigure} 
\subsection{Wires}
A special case of our model is a \emph{wire}, which can be thought of as a
tension-only rod. A wire $ij$ has zero resistance to bending and compression
(i.e., $\sigma^c_{ij} = \sigma^s_{ij} = 0$) and very high resistance to tension
(i.e., $\sigma^t_{ij} >> 0$).
Wires made of strong material such as braided steel can be very thin (near
invisible) while maintaining high strength. 
Our method will allow a mixture of tension-compression-bending rods (e.g.,
wooden dowels) and tension-only wires (steel wires), see \reffig{pterosaurus}.
As special case, we can limit our optimization to consider only wires, resulting
in a hanging optimization (see \reffigs{cupped-sgp}{wire-bird}).
\begin{figure*}[t!]
  \includegraphics[width=\linewidth]{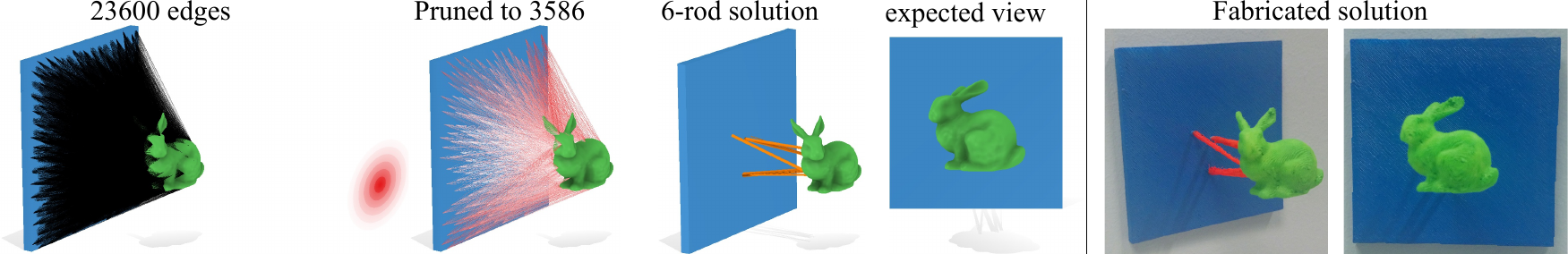}
  \caption{\label{fig:bunny-ground-structure}
  Our method constructs a over-connected ground structure of candidate edges
  (left) then immediately prunes edges that intersect the scene (middle) and
  finally extracts a small number of hidden rods.
  Savings from pruning can produce $10\times$ performance improvements.
  }
  \vspace*{-0.5cm}
\end{figure*}
\subsection{Visibility}
%
% didn't have time
% \review{reviewer 1 says the explanation of the visibility calculation is confusing. suggested a figure.} 
We define the \emph{expected visibility} of a rod as function of the input
viewpoint distribution, occlusions due to the scene, the rod's position and
orientation and its unknown cross-sectional area. For a rod $ij$, its expected
visibility $v_{ij}$ is:
\sarahchange{
\begin{align}
  v_{ij} &= 
    \int_\mathcal{E} p(\e) 
    \int_{C_{ij}}  r(\e,\x) \;d\Omega
    \; d\e,
  \intertext{where}
  r(\e,\x)  &= 
    \begin{cases}
      0 & \text{if the segment $\e\x$ intersects the scene}, \\
      1 & \text{otherwise},
    \end{cases} \notag
\end{align}
}
\begin{wrapfigure}[10]{r}{1.3in}
  % \vspace{-0.5cm}
  \includegraphics[width=\linewidth,trim={1.5mm 0mm 0mm 1.5mm}]{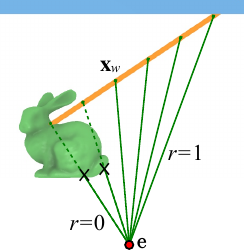} 
  \label{fig:visibility-sketch}
\end{wrapfigure} 
where $\mathcal{E}$ defines the set of viewpoints and $p(\e)$ is the probability
density associated with the point $\e \in \mathcal{E}$, and $C_{ij}$ is the
surface of the cylindrical rod with cross-sectional area $a_{ij}$ connecting
endpoints $\x_i$ and $\x_j$, and $d\Omega$ is the differential solid angle at
the corresponding integration point $\x$ subtended at the viewpoint $\e$.
Measuring visibility according to solid angle correctly matches the intuition
that the same size rod farther away from an observer is less visible.

The outer integral is immediately recognizable as a soft-shadow or area-light source
evaluation common in rendering. We can approximate this well by Monte-Carlo
importance sampling over the viewpoint distribution.
An analytic expression for the inner integral becomes unwieldy, so we instead
opt for a simple approximation based on uniform quadrature, accounting for the
orientation of the rod resulting in foreshortened projection.
Rods are thin relative to the scene and spread of the viewpoint distributions,
therefore we assume visibility to be constant in the normal directions of the
rod.
Our discrete approximation of the expected visibility is thus a double sum over
$n_u$ points sampled according to the input probability density function and
$n_{ij}$ points sampled along the rod:
\begin{align}
  \label{equ:vijapprox}
  v_{ij} &\approx 
    \sqrt{a_{ij}}
    \underbrace{
    \frac{1}{n_u\sqrt{2\pi}}
      \sum_{u=1}^{n_u}
      \cos^{-1}\left(\frac{(\x_i-\e_u)\cdot
      (\x_j-\e_u)}{\|\x_i-\e_u\|\|\x_j-\e_u\|}\right)
        \sum_{w=1}^{n_{ij}}
          r(\e_u,\x_w)}_{g_{ij}}, \notag
\end{align}
where we collect the terms that do not depend on $a_{ij}$ into a single
non-negative scalar per-rod, $g_{ij} \in \R_{\ge 0}$.
In this way, the \emph{squared visibility} of each rod becomes a linear function
of the cross-sectional area:
$ a_{ij} g_{ij}^2$.
Because \emph{wires} are so thin compared to \emph{rods}, we happily set
$g_{ij}=0$ for \emph{wires} and avoid their visibility precomputation.
To generate the $n_{ij}$ samples on edge $ij$, we subdivide the edge until all
segments are less than a given scene-dependent length threshold (e.g., $0.1$ meters 
for the bedroom scene in \reffig{teaser}) and then use the segment barycenters
as samples (typically 10-100 samples per edge).
Segment queries can be computed in parallel.

\subsection{Ground Structure}
The space of physically feasible supporting structures is high-dimensional and a
mixture of discrete variables (e.g., how many rods? connecting between which
objects?) and continuous variables (e.g., where rods attach to each object?
what are the rod thicknesses?).
Navigating this space to find a \emph{globally optimal} solution is difficult.
In response, the ground structure method (e.g., \cite{dorn1964automatic, pedersen1993topology, zegard2015grand3,freund2004}
makes the problem tractable by
rephrasing the problem into \emph{selecting} a discrete subset of support
elements from an intentionally dense yet finite set of candidate elements. This
candidate set is referred to as the ``ground structure.''

In our case, we generate a ground structure of candidate rod and wire elements
by Poisson disk sampling \cite{Yuksel2015} all rigid objects and the support
surface and then connecting all possible pairs of points from different sources
(e.g, for a single rigid object this forms a bipartite graph with the supporting
surface, see \reffig{bunny-ground-structure}).
For each edge in this graph, we label it as a ``rod'' or ``wire'' (and possibly
create duplicate copies so edges appear as both types).
We can discard a \emph{bad} edge $ij$ if its attachment angle is self-penetrating
or too obtuse (by checking if the rod vector dotted with the surface normal is below 
a threshold; $\that_{ij}\cdot \hat{\n}_i < \cos
\theta_\text{max}$), if it intersects objects in the scene (by ray casting), or
if its computed visibility coefficient is exceptionally high ($g_{ij} >
g_\text{max}$).
We refer to the result as the \emph{pruned} ground structure $\GS$.
\begin{figure}[b!]
  \includegraphics[width=\linewidth]{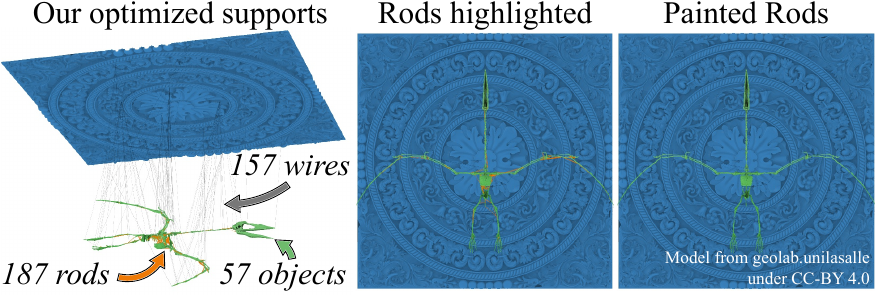}
  \caption{\label{fig:pterosaurus} 
   We show the rods in orange to demonstrate how hidden they are. But we show that the
   rods connecting the smaller parts (middle) can be painted to blend in with the ceiling (right).
  }
\end{figure}

\subsection{Sparse Optimization}
The beauty of the ground structure method is that once the candidate set has
been chosen, selecting the globally optimal subset can be phrased as an
efficient convex optimization, in particular a linear program.
In the classic method, the cost function to be minimized is the total volume of
material spent on the support structure. Since all edge lengths are known once the
candidate set is selected, this cost is a linear function of the yet unknown
edge cross-sectional areas.
It is important that this cost function is the \emph{unsquared} volume, which
can be thought of as the L1-norm of the vector of edge areas (weighted by
edge-lengths), as opposed to the sum of squared per-edge volumes, analogous to
the L2-norm.
The L1-norm is \emph{sparsity inducing} and under mild conditions will agree
with the optimal solution of the selection problem, analogous to the
L0-pseudonorm \sarahchange{ ~\cite{candes2008enhancing,feng2013complementarity}}.
As a result, the vast majority of edges in the solution will have exactly zero
area.

In our case, we augment the total volume cost function with a least-squares
visibility term to penalize choosing highly visible rods. Because our per-edge
visibility measurement in \refequ{vijapprox} is linear in the square-root of the
rod areas, this least-squares energy becomes linear in the areas.

The areas of the rods and wires are the primary unknowns. We introduce auxiliary
variables $c_{ij}$ and $\q_{ij}$ as described in \refsec{rods} to facilitate
writing our force and torque balance constraints (see \refsec{equilibrium}).
These variables are then coupled to the areas via the yield stress inequalities
(see \refequ{yieldb}).

The resulting optimization is a linear program over the pruned ground structure
$\GS$ containing $m$ candidate edges:
\begin{align}
  \mathop{\text{min}}_{\a,\c,\q} & \sum_{ij \in \GS} a_{ij} (\ell_{ij} + \lambda g_{ij}^2) 
  \\
    \text{s.t.} 
                    % force balance on each object
    \label{equ:stbegin}
                    & \sum_{ij | j\in V_k}
                    c_{ij} \hat{\t}_{ij} +
                    {\N}_{ij} \q_{ij}
                    = m_k \g ,
                    \;\forall \; k=1,\dots,K\\
                    % torque balance on each object
                    & \sum_{ij | j\in V_k}
                    (c_{ij} \hat{\t}_{ij} +
                    {\N}_{ij} \q_{ij}) \times (\x_j -
                    \overline{\x}_k )  = 0,
                    \;\forall \; k=1,\dots,K \notag \\
                    & -\sigma_{ij}^c a_{ij} \leq c_{ij} \leq \sigma_{ij}^t
                    a_{ij},
                    \; \forall \; ij \in \GS \\
    \label{equ:stend}
                    & -\sigma_{ij}^s a_{ij} \leq \q_{ij} \leq
                    \sigma_{ij}^s a_{ij},
                    \; \forall \; ij \in \GS \\
                    & a_{ij} \ge 0,
                    \; \forall \; ij \in \GS
\end{align}
where we stack all $a_{ij}$, $c_{ij}$, and $\q_{ij}$ variables into vectors $\a
\in \R^{m}$, $\c \in \R^{m}$, and $\q \in \R^{2m}$, respectively, and we introduce
the user-controllable weighting term $\lambda$ to balance between preference for
volume and visibility minimization. \sarahchange{For all examples shown, we use $\lambda$ = 10,000.}

We opt to replace the second-order cone constraint for linearized bending yields in
\refequ{yieldb} with the simpler coordinate-wise linear inequality in
\refequ{stend}. This can be thought of as a conservative $L_\infty$ approximation,
and albeit coordinate system dependent, does not affect results and admits a
faster linear program than a conic program in our experience.

The linear coefficients in the force/torque balance equations and linear
inequalities (Eqs.~\ref{equ:stbegin}-\ref{equ:stend}) can be collected in large
\emph{sparse} matrices (see \refapp{sparse}). Many efficient solvers exist for
such large sparse linear programs; we use \textsc{Mosek} \cite{mosek}.

A solution is a guaranteed to exist as long as force and torque balance can be
achieved. This could fail to happen for very sparse ground structures (e.g.,
less than six edges per object) or degenerate situations (e.g., all edges are
parallel). Our very dense ground structure (hundreds of thousands of edges)
enjoys the general position of its random providence. We never fail to find a
feasible solution.

\section{Experiments \& Results}
\label{sec:results}

%
%We implemented a prototype of our algorithm in \textsc{Matlab} using
%\textsc{gptoolbox} \cite{gptoolbox} for geometry processing and \textsc{Mosek}
%to solve our linear program.
%
%We use the \textsc{Embree} \cite{embree} ray-tracer as interfaced by
%\textsc{libigl} \cite{libigl} to accelerate our integrated visibility
%pre-computation.
%

% !TEX root = winding.tex
\begin{table}
\centering
\ra{1.2}
\setlength{\tabcolsep}{5.5pt}
\rowcolors{2}{white}{lightbluishgrey}
\setlength\tabcolsep{4pt}
\begin{tabular}{l r r r r r r r r r r r r r r r r r r}
\rowcolor{white}
  Scene               & $K$  & \textsc{full}     & $m$    &  \textsc{time} & $R$ & $W$\\
\midrule
Parade Float        & 1    & 20K    & 10K  &   0.14  & 0 &  12  \\
``Koons'' Display   & 1    & 154K   & 22K  &   4.90  & 2 &  4   \\
Ghost With Tail     & 1    & 11K   & 1K   &  5.01   & 5 &  0   \\
Bunny/Teapot/Rocker & 3    & 150K   & 21K  &   2.04   & 16 & 0  \\
Bedroom             & 13   & 490K   & 46K  &  35.87  & 41 &  35    \\
Zoetrope (1 frame)& 1    & 469K   & 51K  &  33.77 &  4 &  0\\
Pterosaurus         & 57   & 10M & 785K & 608.01    & 187 & 157  \\
\bottomrule
\end{tabular}
  \caption{Timings in seconds (\textsc{time}) and numbers of rods ($R$) and wires ($W$)
  for each result with $K$ objects.
  \textsc{full} and $m$ are edges in the original and pruned ground structures,
  respectively. 
The ``Pterosaurus'' and ``Parade Float'' examples do not include the time for
  computing visibility, as it was not used in the LP objective.
}
\label{tab:timings}
    \vspace*{-6mm}
\end{table}

We implemented our algorithm in \textsc{Matlab} using \textsc{gptoolbox}
\cite{gptoolbox} for geometry processing and \textsc{Mosek} \cite{mosek} to solve the linear
program formulated in Section 3.6. Pre-computation of the integrated visibility,
in our input scene is accelerated using the \textsc{Embree} \cite{embree}
ray-tracer as interfaced by \textsc{libigl} \cite{libigl}. 
%\sarahchange{For all examples shown we used a visibility weight value of $\lambda=10,000$.} 
We report statistics and timings for the results in our paper in \reftab{timings}.  
All times are reported on a MacBook Pro with 3.5 GHz Intel Core i7 and 16GB of RAM.
Visibility pre-computation is computed in parallel, but is still typically the
bottleneck ($\approx 80\%$).

\begin{wrapfigure}[7]{r}{1.6in}
  \hspace{-1.5mm}
  \includegraphics[width=\linewidth,trim=13mm 5mm 3mm 8mm]{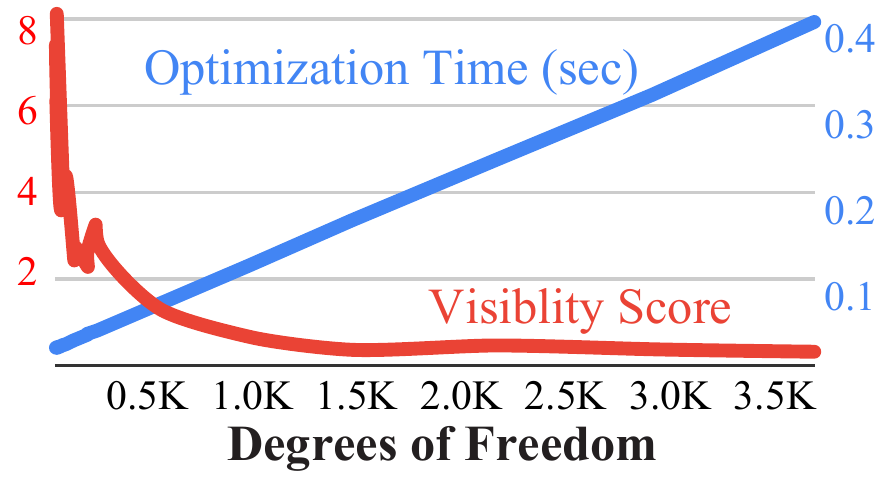}
  \vspace{4mm}
  \end{wrapfigure}
The number of degrees of freedom in the system is the size of the ground
structure which generally scales quadratically in the number of objects
$m = O(K^2)$, typically generated by taking all inter-object pairs over 10-100
Poisson disk sample points on each object. 
\risachange{ The inset graph shows the
effect of increasing the degrees of freedom on \reffig{bunny-teapot-rocker-arm}. While
 optimization time increases linearly with degrees of freedom, 
improvement to the visibility score of the solution reaches a point of diminishing return. }

\begin{figure}
    \includegraphics[width=\linewidth]{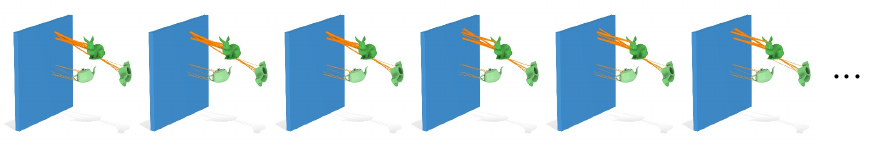}
    \caption{\label{fig:random-sampling}
    Our results depend on a randomly generated ground structure. Changing the random seed 
    affects the precise result, but not qualitatively.
    }
%\end{figure}
%\begin{figure}
    \includegraphics[width=\linewidth]{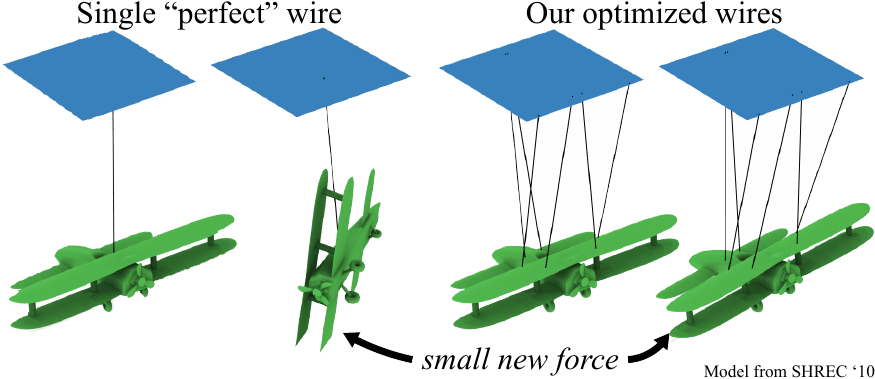}
    \caption{\label{fig:single-wire-failure}
    Applying a small new force to the plane held by a single wire causes undesired behaviour since
    a single wire attachment is not enough to balance the torque. Our method gives a 6-wire solution,
    exactly the number needed to balance force and torque.
    }
    \vspace*{-6mm}
  \end{figure}

Starting with a dense ground structure leads to better qualitative results, but the
exact positions of the samples do not drastically effect the hidden-ness of the result.
\reffig{random-sampling} shows how little the solution changes as a function of the
ground structure sampling. 

Pruning often significantly reduces the ground structure size and consequently, the number of degrees of
freedom (see, e.g., \reffig{bunny-ground-structure}).
Perhaps unsurprisingly, we typically experience a speedup the same ratio of original ground 
structure edges to pruned ground structure edges.
The number of constraints in our optimization is six times the number of
objects $K$.
After pruning, \textsc{Mosek} finds a solution for the above problem configuration within a few minutes.

In our accompanying video, we show animations of results in this paper including
traversals of the viewing distributions to demonstrate the robustness of our
methods ability to hide supports.

\begin{figure*}
  \includegraphics[width=\linewidth]{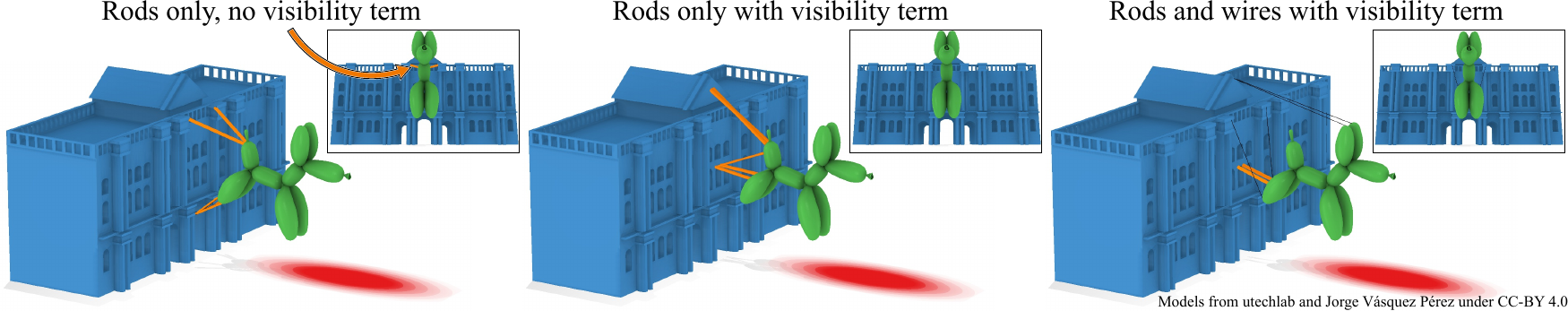}
  \caption{\label{fig:koons}
  Applying the ground structure method to this example of a giant balloon hanging outside of a museum gives
  sufficient rods to support it, but they are visible. Using our visibility term in the optimization yields
  a support structure with rods hidden to the viewpoints. Allowing wires for tension and rods for compression,
  the result is a few thick but invisible rods and thin wires which hold the balloon in place.}
  \vspace*{-5mm}
\end{figure*}

\begin{figure}
  \includegraphics[width=\linewidth]{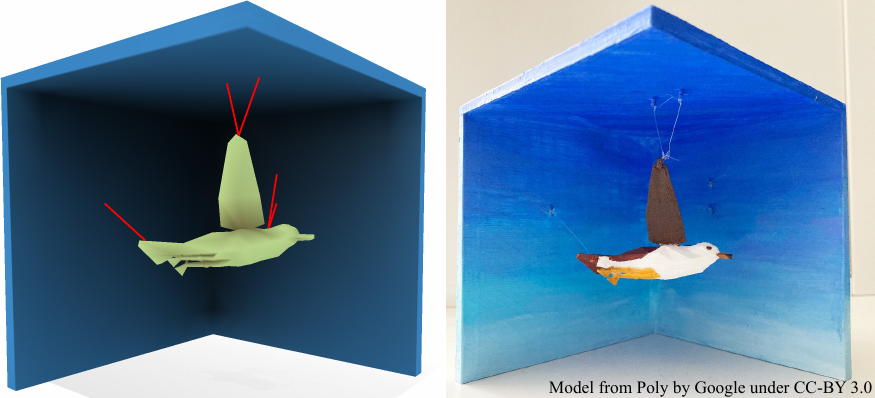}
  \caption{\label{fig:wire-bird} \sarahchange{For a wire-only solution, we can save time by forgoing
  the visibility computation. Using fishing wire, we support the seagull
  in mid-air invisibly.}}
  \vspace*{-8mm}
\end{figure}

Levitating 3D objects has a wide range of applications including scientific
visualization, film and theater set design, home decor, anamorphic 3D art
installations, as well as objects for zoetropes and 3D stop-motion animation.
Each application has specific design requirements, and our algorithm is designed
to enable the exploration of a number of aesthetic and structural parameters and
design choices, which significantly impact the resulting solution. We elaborate
on some of these design use cases.

\begin{figure}
  \includegraphics[width=\linewidth]{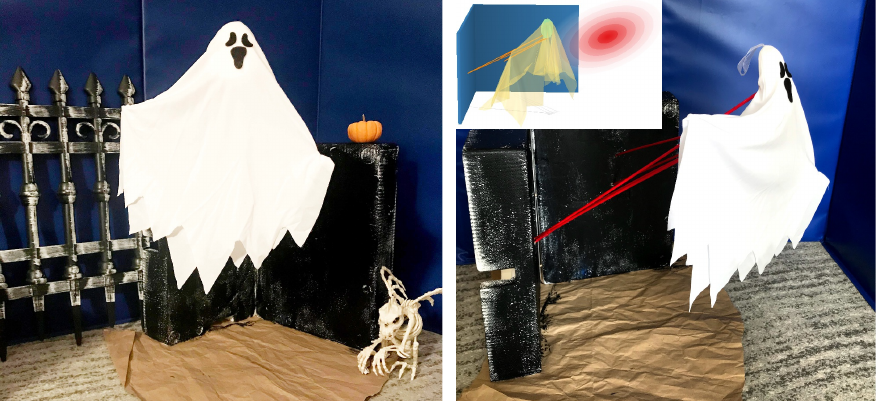}
  \caption{\label{fig:ghost}
  The model used for support attachments does not have to be the same one used for visibility. 
  The ghost's head (green) has attached supports, while its tail (yellow) hides them.
  }
  \vspace*{-6mm}
\end{figure}

Both scientific exhibits (see, e.g., \reffigs{teaser}{ubc-whale}) and illusory art
installations (see \reffig{ghost}) require an unobscured view
of the levitating objects.
While it is feasible for designers to hand-craft
support structures from a single fixed viewpoint, the interplay between
visibility and structural stability is quite complex for mutli-view
distributions.
In \reffig{teaser}, we show the ability of our algorithm to adapt its
optimal solution to multiple viewpoint distributions.

%% Alec: I don't understand this paragraph.
%Objects can be designed to levitate in sizes ranging from a tiny doll house for
%zoetropes and stop motion, to the face of a large building in a public square
%\reffig{fig:backflip-zoetrope,koons}. In general, the diameter of the rods and
%wires used do not scale in proportion with the scale of the scene, and it would
%be easy to add a maximum rod cross-section $a_{ij}<= maxR$ to Equation 12. Large
%scenes may also need to account for other environmental forces like wind when
%optimizing the support structure.

\begin{wrapfigure}[11]{r}{1.25in}
  \includegraphics[width=\linewidth,trim=6mm 0mm 0mm 2mm]{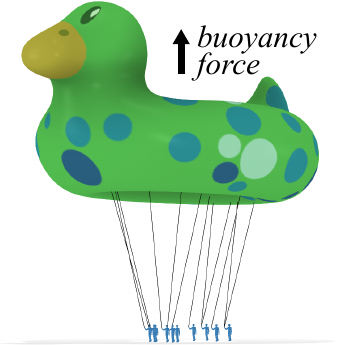}
    %\caption{\label{fig:parade-balloon} }
  \end{wrapfigure} 
Our algorithm is able to holistically optimize the support structure using a mix
of rods and wires. We color our rods bright orange for evaluation in this paper,
but in practice they can be further camouflaged by matching their appearance to
the background or scene objects (see \reffigs{pterosaurus}{sheep}).
The choice of using a rod or wire is both aesthetic (as determined by a user)
and functional. For example, supporting a levitating object with a wire would
require a potential attachment points on the fixed surface or other levitating
objects, to be vertically higher than the given object (see, e.g.,
\reffig{cupped-sgp}). 
The inset figure shows a parade float suspended by optimized wires (the net
force pointing upward due to buoyancy).
Previous methods have considered hanging objects \cite{PrevostWLS13} or more
generally mobiles \cite{Mobiles} by placing a single support ``perfectly''
placed in alignment above the center of mass. While this strategy requires the
fewest supports, it is an unstable solution (see \reffig{single-wire-failure}).
Our method relies on random sampling of points in general position, typically
producing multiple wires per hanging object, but resulting in a more stable
configuration. \risachange{Thus, in practice, we've found both our 3D printed and assembled 
results to be quite resilient to outside forces (including those from falls, heavy winds,
 and moving from one place to another).}

Mounting objects off the side of a support such as a wall is best achieved with
a mixture of wires and rods. 
\reffig{koons} shows a giant promotional display suspended in front of a
contemporary art museum.
We provide a symmetric dense ground structure and our optimization naturally
finds a symmetric sparse solution.

The pterosaurus in \reffig{pterosaurus} has 57 separate bones and requires a
complex support structure, acting as a stress test on our optimization.
In practice, skeleton displays often pre-plaster-fuse bones to reduce the number
of pieces (see spine of whale in \reffig{ubc-whale}).

\begin{figure}
  \includegraphics[width=\linewidth]{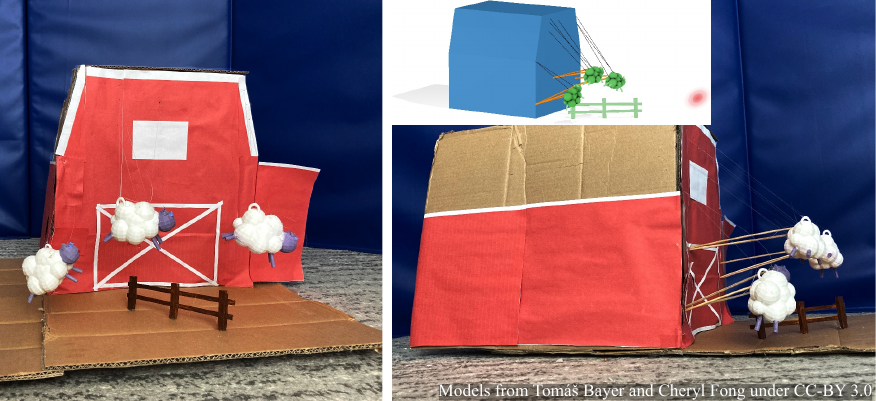}
  \caption{\label{fig:sheep}
  \sarahchange{By maintaining separate graphs for rods and wires, we can use differing visibility
  weights and yield stresses based on what materials are going to be used in fabrication. 
  Our system can wisely select which edges should be wires vs rods.
  }
  % \sarah{should i/ do i need to say somewhere that i've rounded up the radii for assembly?}
  }
  \vspace*{-2mm}
\end{figure}

% This example should come last
\begin{figure}
  \includegraphics[width=\linewidth]{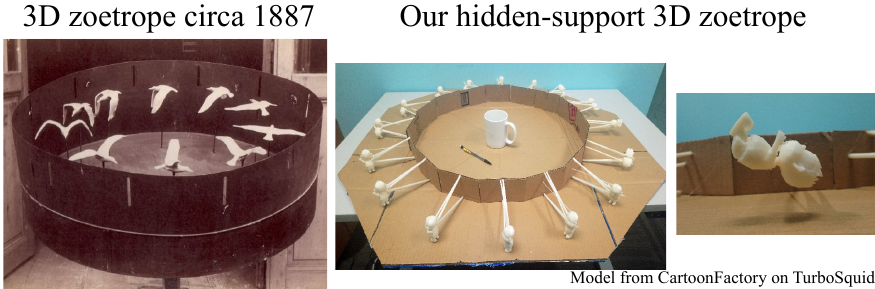}
  \caption{\label{fig:backflip-zoetrope} 3D zoetropes are an old idea, but hiding
    supports for flying objects is still challenging. We incorporate the centripetal
    force due to spinning and hide supports behind a backflipping boy zoetrope.
    See accompanying video at 3m20s.}
    % \sarah{am going into the lab to fix photos}}
    % 
    \vspace*{-6mm}
\end{figure}
The idea of stop-motion animation and 3D zoetropes is over a century old
\cite{marey1890physiologie}, with
modern examples including ``Feral Fount'' by Gregory Barsamian
at the Museum of the Moving Image in Queens and the \emph{Toy Story}
zoetrope featured in Pixar's Museum Exhibit.
The portrayal of levitating objects in this medium is particularly challenging. 
We demonstrate a prototypical result of a backflipping boy in
\reffig{backflip-zoetrope} by hiding supporting rods out of sight.
For this example to be structurally stable both at rest and while spinning, we
first find the optimal set of rods for each frame under gravity and then re-run
the linear program on just these rods subject to centripetal forces. The final
rod thickness are the maximum over the two solves.
Incorporating more elaborate multi-load handling (cf. \cite{freund2004}) is
left as future work.

% \begin{figure}[t!]
%   \includegraphics[width=\linewidth]{figures/bunny-fabrication}
%   \caption{\label{fig:3dprint-bunny}
%   \alec{Is this bunny the same as the one in \reffig{bunny-ground-structure}? Combine or at least
%   don't repeat the same/slightly different digital solution.}
%   \sarahchange{In this 3D print, we show that our algorithm does indeed
%   produce sufficiently strong rods to hold up the bunny.}}
% \end{figure}

% \sarah{didn't have time for this. compare our rod and wire combinations to FEA}

\section{Limitations \& Future Work}
\begin{figure}
    \includegraphics[width=\linewidth]{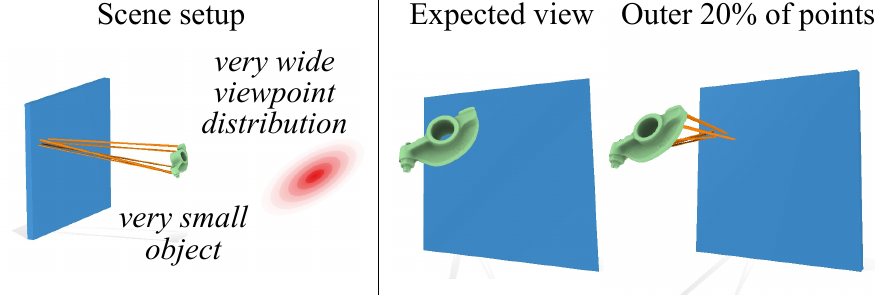}
    \caption{\label{fig:failure-case}
    In the case of a very wide viewpoint distribution and a small or thin object,
    there will most likely be viewpoints from which the supports are visible. The
    rightmost figure shows the scene from a viewpoint on the outer 20\% of the distribution.}
    \vspace*{-4mm}
  \end{figure}

Our rod model includes linearized tension, compression, and bending forces.
Like many past methods, we do not handle the self-weight of the rods by assuming
that the force of gravity is much larger than the force of the rods on
themselves.
This is a trivial addition of gravity forces on each rod proportional to their
length.
Ground structure methods may produce solutions where thickened rods intersect; ours is no exception.
\sarahchange{Edges which nearly overlap with each other appear in the original
ground structure and therefore may be selected as rods in the solution.
However, this has not caused any fabrication problems in practice.}
Previous methods have considered penalty terms or post-pressing to deal with
intersecting (e.g., \cite{JiangTSW17}).
Wire-wire intersections are extremely unlikely due to the very thin nature of
wires. 
% \risachange{add explanation of double counting here}
%
Our visibility model considers direct line of sight, but not other cues such as
reflections or shadows.
Transparency of objects is not accounted for.
Depending on the setup of the scene, there may not be a solution invisible to every
viewpoint (e.g., \reffig{failure-case}). 
Since we model physical validity as a hard constraint, we are still able to find a solution, albeit a
visible one.

% Sometimes the output of our algorithm contains rods that join up at the same attachment points. 
% A reasonable work-around might be to, once we've gotten an initial truss, perturb the attachment points
% slightly to 

The precise solution depends on the initial ground structure. In general, denser
ground structures produce higher quality solutions --- both in terms of total
structure volume and hidden-ness --- with diminishing returns.
Rod areas are directly proportional to stress limits, so acurate fabrication
relies on accurate (or at least conservative) material measurement.
%
% In fact, fabrication of our digital results would also be an exciting next step.

% \sarah{we do have fabrication now. removed paragraph from future work}

% Fabrication may be within reach utilizing recent wire-frame
% methods \cite{KovacsSWCMMYBSK17, ChidambaramZSER19, Jacobson19} or via 3D
% printing.
% %
% A voxel-based topology optimization formulation of our problem may be better
% suited for 3D printing or milling pipelines.
% %
% It may be possible to forgo fabricating a literal support structure entirely by
% exploring magnetic levitation.

%Fabrication is le
%on assembling these types of structures . Interesting future work
%might be inspired by the notion that curved edges could possibly be better hidden in occluded areas of the scene since
%these regions are not necessarily large and convex.
%But these are more difficult to optimize over and to fabricate. To this end, voxel-based topology optimization might be better 
%suited for advanced manufacturing like 3D printing and milling. It would also be extremely fun to fabricate these scenes by
%levitating objects using magnets \cite{DBLP:journals/tog/KimPH18}.

Our algorithm assumes that the input is a well-crafted scene to begin with and
leaves it perfectly as inputted.
The creative design process for these scenes is itself non-trivial.
In the future, we are interested in pursuing an interactive design tool which would provide hints to
increase occlusion by applying simple transformations (translations, rotations
and scales) to the objects in the scene or even provide automatic layout
optimizations given the objects and the viewpoints.

%In summary, we propose a novel computational problem of creating a physically
%stable and hidden structure of rods and wires, to support an input spatial
%arrangement of objects, so they appear to levitate from a given distribution of
%viewpoints. We solve this problem efficiently using a ground structure-like
%method with forces and design objectives formulated as the optimal solution to a
%linear program. We evaluate our approach with compelling results on inputs
%that exemplify a variety of scientific and artistic applications.

We model the problem of hidden supports as an efficient linear program that
leverages fast ray-casting from computer graphics.
We see an exciting future in combining techniques from rendering and geometry
processing with structural optimization in architecture and engineering.
We hope this combination of appearance-driven design will be beneficial to
scientific and artistic endeavours. 

\section{Acknowledgements}
This research is funded by 
New Frontiers of Research Fund (NFRFE–201),
NSERC Discovery (RGPIN2017–05235, RGPAS–2017–507938), 
the Ontario Early Research Award program, 
the Canada Research Chairs Program, 
the Fields Centre for Quantitative Analysis and Modelling and gifts by Adobe Systems, 
Autodesk and MESH Inc.
We would like to thank
Michael Tao,
Silvia Sell\'{a}n and
Rinat Abdrashitov
for proofreading;
John Hancock for IT and fabrication hardware support;
the anonymous reviewers for their helpful comments.

\printbibliography

% why can't I use \appendix ?

\section*{Appendix: Matrix Form}
\label{app:sparse}
Solvers like \textsc{Mosek} \cite{mosek} expect the problem to be provided in
matrix form. We spell out the coefficients of the relevant sparse matrices
implementing the linear program in \refequ{stbegin}.

% \review{i and j are used to select different numbers of rows/columns in each 
% matrix defined here. we use rod ij earlier in the paper to mean a rod connecting points
% i and j. but here i means a rod and j is one of its vertices. we should either explain this
% or change it.}\\
For our pruned ground structure  with $m$ candidate edges connecting $N$
vertices, introduce a unit-less sparse matrix $\C \in \R^{3 N \times m}$ where:
\vspace*{-4mm}
\sarahchange{
  \begin{equation}
\C_{jl} = \begin{cases}
  \hphantom{-}\hat{\t}_{ij} & \text{ if rod $ij$ points toward $\x_j$ } \\
  -\hat{\t}_{ij} & \text{ if rod $ij$ points away from $\x_j$ } \\
  \hphantom{-}0 & \text{ otherwise.}
\end{cases}
\end{equation}}
\vspace*{-4mm}

Here, $j$ is used to index the 3 rows that correspond to the vertex $\x_j$ and
$l$ is used to index the column for rod $ij$.

Introduce a unit-less sparse matrix $\Q \in \R^{3N \times 2m}$ where
\vspace*{-1mm}
\sarahchange{
  \begin{equation}
\Q_{jl} = \begin{cases}
  \hphantom{-}\hat{\N}_{ij} & \text{ if rod $ij$ points toward $\x_j$ } \\
  -\hat{\N}_{ij} & \text{ if rod $ij$ points away from $\x_j$ } \\
  \hphantom{-}0 & \text{ otherwise.}
\end{cases}
\end{equation}}
\vspace*{-4mm}

Introduce a sparse unit-less selection matrix $\S \in \R^{3K\times 3N}$, where
\vspace*{-4mm}
\sarahchange{
  \begin{equation}
\S_{kj} = \begin{cases}
I_3 &  \text{ if vertex $\x_j$ lies on object $k$} \\
0 & \text{ otherwise}
\end{cases}
\end{equation}
}
\vspace*{-5mm}

Introduce a sparse cross-product matrix $\D \in \R^{3K\times 3N}$ with units
meters, where
\vspace*{-1mm}
\sarahchange{
  \begin{equation}
\D_{kj} = \begin{cases}
[\x_j -\overline{\x}_k]_\times &  \text{ if vertex $\x_j$ lies on object $k$} \\
0 & \text{ otherwise}
\end{cases}
\end{equation}
}
where
\vspace*{-4mm}
\sarahchange{
\begin{equation}
    [\d]_\times = \begin{bmatrix}
    0 & -\d_3 & \d_2 \\
    \d_3 & 0 & -\d_1 \\
    -\d_2 & \d_1 & 0
    \end{bmatrix} \in \R^{3\times3}
\end{equation}
}
\vspace*{-2mm}

% \review{changed G and g to D and d because g in the next part means gravity.}
Finally, the full linear program in matrix form may be written 
% \review{need to 
% have v element-wise squared in the objective.
% changed $v$ to $g$ because $v = a*g$}:
\vspace*{-1mm}
\begin{align}
    \mathop{\text{min}}\limits_{\a, \c, \q} & \quad {(\ell + \lambda g')}^\top \a \\
    \text{subject to } & \quad \begin{bmatrix}
    \Zero & \S \C & \S \Q \\
    \Zero & \D \C & \D \Q
    \end{bmatrix}
    \begin{bmatrix}
    \a \\
    \c \\
    \q
    \end{bmatrix}
    =
    \begin{bmatrix}
    \m \otimes \g \\ 0
    \end{bmatrix}
    \\
    \text{and } & \quad -\sigma_t \a_l \leq \c_l \leq  \a_l\sigma_c , \forall l  \\
    \text{and } & \quad -\sigma_s \a_l \leq \q_l \leq \a_l\sigma_s
    , \forall l.
\end{align}

\sarahchange{where 
$g'_l = g_l^2, \hspace{1mm} \forall l$ since the squared visibility is linear in the cross-sectional areas of each rod.}

\sarahchange{$\m \otimes \g$ denotes the Kronecker product of the $\m \in \R^K$ stacked
vector of object masses and the $\g \in \R^{3 \times 1}$ gravity vector.}

\end{document}